\documentclass[prb,twocolumn,groupedaddress]{revtex4}
\usepackage{amssymb}
\usepackage{amsmath}
\usepackage{amsfonts}
\usepackage{graphicx}
\usepackage{bm}
\newcommand* {\vek}[1]{{\ensuremath{\bm{\mathrm{#1}}}}}
\newcommand* {\vekc}[1]{{\ensuremath{\bm{\mathcal{#1}}}}}
\newcommand* {\kk}{\vek{k}}
\newcommand* {\rr}{\vek{r}}
\newcommand* {\frack}[2]{{\Ts\frac{#1}{#2}}}

\newcommand* {\Ts}{\textstyle}
\newcommand* {\bra}[1]{\ensuremath{\langle {#1} |}}
\newcommand* {\ket}[1]{\ensuremath{| {#1} \rangle}}
\usepackage{array}
\newcolumntype {s}[1]{@{\hspace{#1}}} 

\newcommand* {\kcomp}{\kappa}
\newcommand* {\kvek}{\bm{\kcomp}}
\newcommand* {\tvek}[2][c]{\left( \begin{array}{s{0.15em}#1s{0.15em}}
     #2\end{array} \right)}
\newcommand* {\strain}{\epsilon}
\newcommand* {\Strain}{\vek{\epsilon}}
\newcommand* {\Ee}{\mathcal{E}}
\newcommand* {\koeff}[3]{#1^{#2}_{#3}}

\newenvironment{textmath}{$\displaystyle}{$}

\newcommand{\allowed}[1]{\bm{#1}}
\newcommand{\forbidden}[1]{#1}

\begin{document}

\title{Invariant expansion for the trigonal band structure of graphene}

\author{R. Winkler}
\affiliation{Department of Physics, Northern Illinois University,
DeKalb, Illinois 60115, USA}
\affiliation{Materials Science Division, Argonne National
Laboratory, Argonne, Illinois 60439, USA}

\author{U. Z\"ulicke}
\affiliation{Institute of Fundamental Sciences and MacDiarmid Institute
for Advanced Materials and Nanotechnology, Massey University,
Manawatu Campus, Private Bag 11~222, Palmerston North, New Zealand}
\affiliation{Centre for Theoretical Chemistry and Physics, Massey University,
Albany Campus, Private Bag 102904, North Shore MSC, Auckland 0745,
New Zealand}

\date{14 December 2010}

\begin{abstract}
  We present a symmetry analysis of the trigonal band structure in
  graphene, elucidating the transformational properties of the
  underlying basis functions and the crucial role of time-reversal
  invariance. Group theory is used to derive an invariant expansion
  of the Hamiltonian for electron states near the $\vek{K}$ points
  of the graphene Brillouin zone. Besides yielding the
  characteristic $k$-linear dispersion and higher-oder corrections
  to it, this approach enables the systematic incorporation of all
  terms arising from external electric and magnetic fields, strain,
  and spin-orbit coupling up to any desired order. Several new
  contributions are found, in addition to reproducing results
  obtained previously within tight-binding calculations. Physical
  ramifications of these new terms are discussed.
\end{abstract}


\maketitle

\section{Introduction}

\begin{figure}[tbp]
  \includegraphics[width=1.0\columnwidth]{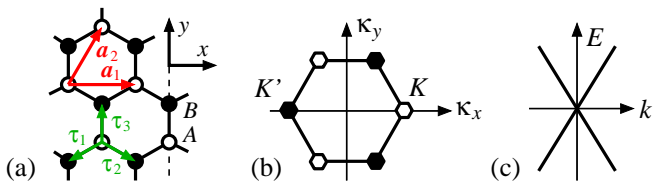}
  \caption{\label{fig:lattice} (Color online) (a) Honeycomb
   structure of graphene. Atoms in sublattice $A$ ($B$) are marked
   with open (closed) circles. (b) Brillouin zone and its two
   inequivalent corner points $\vek{K}$ and $\vek{K}'$. The
   remaining corners are related with $\vek{K}$ or $\vek{K}'$ by
   reciprocal lattice vectors. (c) Dispersion $E(k)$ near the
   $\vek{K}$ point. We have $\kk \equiv \kvek - \vek{K}$.}
\end{figure}

In recent years, tremendous interest has been focused on studying
single layers of graphene, \cite{gei07, bee08, cas09} following the
first experimental realization of this material. \cite{nov04,
 nov05b} To a large extent, these continuing efforts are motivated
by the unique band structure of graphene in the vicinity of the
Fermi edge. In most semiconductors, the band edges are characterized
by a quadratic dispersion, with $k$-linear corrections possible only
in inversion-asymmetric materials due to spin-orbit (SO) coupling.
\cite{bir74} In contrast, for graphene, the dispersion $E(\kk)$ of
the uppermost valence band and the lowest conduction band is {\em
 dominated\/} by $k$-linear terms, \cite{wal47} $E(\kk) \approx
\hbar v k$ with Fermi velocity $v$. These bands touch at the points
$\vek{K}$ and $\vek{K}'$ at the edge of the Brillouin zone [Fig.\
\ref{fig:lattice}(b)] so that the resulting energy surfaces resemble
those of free massless fermions described by the Dirac equation
[Fig.\ \ref{fig:lattice}(c)]. The apparent analogies between a
solid-state system and relativistic quantum mechanics have greatly
stimulated the interest in graphene. \cite{gei07, bee08, cas09,
 sem84, div84, hal88}.

Both in early work \cite{wal47} as well as in more recent
pu\-bli\-ca\-tions \cite{sem84, div84, hal88, sai98, cas09} the
electronic properties of graphene have largely been explored using
tight-binding calculations. Tight-binding models also provide the
usual starting point for the derivation of simplified, effective
Hamiltonians \cite{div84, cas09} to describe the $k$-linear
dispersion in the vicinity of the Fermi energy [Fig.\
\ref{fig:lattice}(c)]. Alternatively, first-principles methods have
also been applied to graphene. \cite{pai70} The importance of group
theory for the characterization of the graphene band structure was
already recognized in early work. \cite{car53, lom55, slo55, slo58,
 bas67} Here we employ a group theoretical approach to graphene
\cite{bir74} that was pioneered for the study of bulk semiconductors
\cite{lut56} and, more recently, has proved to be useful also for
the systematic investigation of band-structure effects in
low-dimensional systems. \cite{win03} Using the theory of invariants
\cite{bir74} we find the Hamiltonian describing electronic degrees
of freedom near the $\vek{K}$ and $\vek{K}'$ points in graphene in
terms of a systematic expansion in orders of wave-vector difference
$\kk$ ($\equiv \kvek - \vek{K}$ or $\kvek - \vek{K}'$, respectively)
from these special points and various external perturbations. Both
the sublattice-related (orbital) pseudospin and the intrinsic spin
of quasiparticles in graphene are accounted for within this scheme,
and the fundamental difference in the origin of these two degrees of
freedom is reflected in the transformational properties of the
Bloch-state basis functions. In principle, our analysis can be used
to construct all allowed terms in the Hamiltonian up to any desired
order; but we limit the present discussion to all contributions up
to second order as well as selected terms up to third order. Several
new terms are found that have not been obtained previously.

The important role played by discrete symmetries in protecting a
$k$-linear dispersion in graphene was pointed out in
Ref.~\onlinecite{man07}. Here we discuss in detail the special
significance of time-reversal invariance for the trigonal band
structure of graphene. It provides an additional criterion that is
satisfied only by a subset of all terms allowed by the spatial
symmetry of the $\vek{K}$- and $\vek{K}'$-point Bloch functions.

The remainder of this article is organized as follows. To introduce
our phase conventions and other relevant background information, the
following Sec.~\ref{sec:tb} provides a summary of the
tight-binding description for a graphene sheet. The symmetry
analysis for this system is performed in Sec.~\ref{sec:sym}, with
the role of time-reversal invariance highlighted. A discussion of
the terms obtained within the symmetry analysis is given in
Sec.~\ref{sec:disc} before we present our conclusions in
Sec.~\ref{sec:conc}.

\section{Review of Tight-Binding Analysis}
\label{sec:tb}

The honeycomb structure of graphene is sketched in
Fig.~\ref{fig:lattice}(a). For definiteness, we use the basis vectors in
real space
\begin{equation}
  \label{eq:basis-vec}
  \vek{a}_1 = a \tvek{1 \\ 0} ,
  \qquad
  \vek{a}_2 = a \tvek{1/2 \\ \sqrt{3}/2} ,
\end{equation}
with lattice constant $a$. 
The basis vectors in reciprocal space become
\begin{equation}
  \label{eq:basis-vec-r}
  \vek{b}_1 = \frac{2\pi}{a} \tvek{1 \\ -1/\sqrt{3}} ,
  \qquad
  \vek{b}_2 = \frac{2\pi}{a} \tvek{0 \\ 2/\sqrt{3}} ,
\end{equation}
and the two inequivalent corner points of the Brillouin zone are
\begin{equation}
  \label{eq:Kpoint}
  \vek{K}  = \frac{2\pi}{a} \tvek{2/3 \\ 0}, \quad
  \vek{K}' = \frac{2\pi}{a} \tvek{-2/3 \\ 0}.
\end{equation}

We consider a tight-binding Hamiltonian for the graphene $\pi$ bonds
formed by the carbon $p_z$ orbitals, taking into account
nearest-neighbor and second-nearest-neighbor interactions. For a
given atom in the honeycomb structure, the vectors connecting
nearest-neighbor atoms are ($j=1,2,3$)
\begin{equation}
  \label{eq:nearest-1}
  \vek{\tau}^{(j)}_1 = \mathcal{R} (2j\pi/3) \, \vek{\tau}^{(3)}_1,\quad
  \vek{\tau}^{(3)}_1 = \tvek{a \\ 1/\sqrt{3}} ,
\end{equation}
where $\mathcal{R}(\phi)$ denotes a two-dimensional (2D) rotation by
the angle $\phi$. Similarly, we get the vectors connecting
second-nearest-neighbor atoms ($j=1,\ldots,6$)
\begin{equation}
  \label{eq:nearest-2}
  \vek{\tau}^{(j)}_2 = \mathcal{R} (j\,\pi/3) \, \vek{a}_1.
\end{equation}
Then the tight-binding Hamiltonian becomes \cite{wal47, sai98}
\begin{equation}
  \label{eq:tb-ham}
  \tilde{\mathcal{H}} (\kvek) = 
  \tvek[cc]{\tilde{\epsilon}_{2p} + t_2 f_2 (\kvek) & t_1 \, f_1(\kvek) \\
   t_1 \, f_1^\ast (\kvek) & \tilde{\epsilon}_{2p} + f_2 (\kvek)} ,
\end{equation}
where $\tilde{\epsilon}_{2p}$ is the site energy of the $p_z$ orbitals,
$t_l$ are the transfer integrals for $l$th-nearest neighbors, and the
functions $f_l (\kvek)$ are given by
\begin{equation}
  \label{eq:tb-phase}
  f_l (\kvek) \equiv \sum_j e^{i\kvek\cdot\vek{\tau}^{(j)}_l} .
\end{equation}
The particular geometry (\ref{eq:nearest-1}) gives for $f_1 (\kvek)$
\begin{equation}
  \label{eq:tb-fun-1}
  f_1 (\kvek) = e^{i\kcomp_y a / \sqrt{3}}
  + 2e^{-i\kcomp_y a/2\sqrt{3}} \cos (\kcomp_x a /2) ,
\end{equation}
and we have the relation
\begin{equation}
  \label{eq:tb-fun-2}
  f_2 (\kvek) = |f_1 (\kvek)|^2 -3 .
\end{equation}
Thus it is possible to rewrite the Hamiltonian (\ref{eq:tb-ham})
such that it only depends on the function $f \equiv f_1$,
\begin{equation}
  \label{eq:tb-hamp}
  \mathcal{H} (\kvek) =
  \tvek[cc]{\epsilon_{2p} + t_2 |f|^2 & t_1 f \\
   t_1 f^\ast & \epsilon_{2p} + t_2 |f|^2} ,
\end{equation}
where $\epsilon_{2p} = \tilde{\epsilon}_{2p} - 3 t_2$. We note in
passing that, alternatively, we could have considered a
tight-binding model that takes into account only nearest-neighbor
transfer integrals as well as nearest-neighbor overlap integrals.
\cite{sai98} This model gives qualitatively similar results as the
Hamiltonian (\ref{eq:tb-hamp}). However, extensions of this
alternative tight-binding model, e.g., to include spin-orbit
effects, are hindered by the fact that the matrix of overlap
integrals results in a generalized eigenvalue problem. \cite{sai98}

The tight-binding wave functions corresponding to the Hamiltonian
(\ref{eq:tb-hamp}) are given by
\begin{equation}
  \label{eq:tb-wf}
  \Psi_{\kvek n} (\rr) = \sum_{\lambda=A,B}
  \psi_{\lambda n} (\kvek) \: \Phi_{\kvek \lambda} (\rr) ,
\end{equation}
where $\psi_{\lambda n} (\kvek)$ are $\kvek$-dependent expansion
coefficients for band $n$, and the corresponding basis functions
(that are Bloch functions) are
\begin{equation}
  \label{eq:tb-basis-wf}
  \Phi_{\kvek \lambda} (\rr)
  = \frac{e^{i\kvek\cdot\rr}}{\sqrt{N}} \sum_{\vek{R}_\lambda}
  e^{-i\kvek\cdot (\rr - \vek{R}_\lambda)} \phi_\pi (\rr - \vek{R}_\lambda) .
\end{equation}
Here $\phi_\pi (\rr)$ denote the carbon $\pi$ orbitals, and the sum
runs over the atomic positions $\vek{R}_\lambda$ in sublattice
$\lambda$ [see Fig.~\ref{fig:lattice}(a)].

Diagonalization of the Hamiltonian (\ref{eq:tb-hamp}) yields the
energy dispersion ($n = \pm$)
\begin{equation}
  \label{eq:tb-disp}
  E_n (\kvek) = \epsilon_{2p} + n\, t_1\, |f(\kvek)| + t_2 |f(\kvek)|^2 ,
\end{equation}
with eigenfunctions $\psi_n (\kvek) = (\psi_{A n}, \psi_{B n})$
given by the expressions (valid for $\kvek \ne \vek{K}, \vek{K}'$)
\begin{equation}
  \label{eq:tb-eigenfun}
  \psi_+ (\kvek) = \frac{1}{\sqrt{2}} \tvek{1\\ \hat{f}^\ast (\kvek)}, \quad
  \psi_- (\kvek) = \frac{1}{\sqrt{2}} \tvek{-\hat{f} (\kvek) \\ 1} ,
\end{equation}
where $\hat{f}(\kvek) \equiv f/|f|$. For $t_2=0$, the spectrum
(\ref{eq:tb-disp}) is symmetric around $\epsilon_{2p}$, but this
electron-hole symmetry is broken when $t_2 \ne 0$.

The two bands $\pm$ touch at the $\vek{K}$ and $\vek{K}'$ points of
the Brillouin zone where $f=0$. We can expand $\mathcal{H} (\kvek)$
around $\vek{K}$ ($\kvek \equiv \vek{K} + \kk$), which yields up to
second order in $k$
\begin{equation}
  \label{eq:sham} \hspace{-0.2em}\arraycolsep0.2em
  \begin{array}[b]{rcl}
  \mathcal{H}_{55}^\vek{K} (\kk) & = & \koeff{a}{55}{10} \openone
  + \koeff{a}{55}{61} (k_x \sigma_x + k_y \sigma_y)
  + \koeff{a}{55}{11} (k_x^2 + k_y^2) \openone \\[1ex]
  & & {} + \koeff{a}{55}{62} [(k_y^2 - k_x^2) \sigma_x + 2 k_x k_y \sigma_y]
  + \mathcal{O} (k^3) \, ,
\end{array}\hspace{-1.5em}
\end{equation}
with Pauli matrices $\sigma_j$ and
\begin{subequations}
   \label{eq:coeff}
  \begin{eqnarray}
    \koeff{a}{55}{10} & = & \epsilon_{2p}, \\
    \koeff{a}{55}{61} & = & - \frack{\sqrt{3}}{2} a t_1, \\
    \koeff{a}{55}{11} & = & \frack{3}{4} t_2 a^2, \\
    \koeff{a}{55}{62} & = & - \frack{1}{8} t_1 a^2.
  \end{eqnarray}
\end{subequations}
Our notation will become clear in Sec.~\ref{sec:sym}. The dispersion
becomes
\begin{equation}
  \label{eq:disp}
  E_\pm (\kk) = \koeff{a}{55}{10} \pm \koeff{a}{55}{61} k
  + \koeff{a}{55}{11} k^2
  \pm \koeff{a}{55}{62} k_x (3k_y^2 - k_x^2) / k
  + \mathcal{O} (k^3) .
\end{equation}
It follows immediately from Eq.\ (\ref{eq:tb-eigenfun}) that the
corresponding basis functions describing the two-fold degeneracy at
the $\vek{K}$ point can be chosen to be nonzero only on sublattice
$\lambda = A$ or $B$,
\begin{equation}
  \label{eq:K-wf}
  \Psi_{\vek{K} \lambda} (\rr) = \Phi_{\vek{K}\lambda} (\rr)
  = \frac{e^{i\vek{K}\cdot\rr}}{\sqrt{N}} \sum_{\vek{R}_\lambda}
  e^{-i\vek{K} \cdot (\rr - \vek{R}_\lambda)}
  \phi_\pi (\rr - \vek{R}_\lambda) .
\end{equation}

We obtain the Hamiltonian $\mathcal{H}_{55}^{\vek{K}'} (\kk)$ for
$\vek{K}'$ from $\mathcal{H}_{55}^{\vek{K}} (\kk)$ by applying any
symmetry element $R$ in the point group $D_{6h}$ of the honeycomb
structure that maps $\vek{K}$ onto $\vek{K}'$,
\begin{equation}
  \label{eq:K-Kp}
  \mathcal{H}_{55}^{\vek{K}'} (\kk) = \mathcal{D} (R) \,
  \mathcal{H}_{55}^{\vek{K}} (R^{-1} \kk) \, \mathcal{D}^{-1} (R) .
\end{equation}
Here the matrices $\mathcal{D} (R)$ map the basis functions at
$\vek{K}$ on the basis functions at $\vek{K}'$
\begin{equation}
  \label{eq:K-Kp-wf}
  \Psi_{\vek{K}' \lambda} (\rr) = \sum_{\lambda'}
  \mathcal{D}_{\lambda\lambda'} (R) \,
  \Psi_{\vek{K} \lambda'} (R^{-1}\rr) .
\end{equation}
Choosing for $\vek{K}'$ the basis functions $\Psi_{\vek{K}' \lambda}
(\rr) = \Phi_{\vek{K}'\lambda} (\rr)$ and assuming that, for both
$\vek{K}$ and $\vek{K}'$, these basis functions are ordered as
$(A,B)$, this transformation becomes particularly simple if we consider
the reflection $R=R_y$ at a perpendicular plane that includes the $y$
axis (see Fig.~\ref{fig:lattice}). The reflection $R_y$ preserves
the sublattices $\lambda$, i.e., $\mathcal{D}(R_y) = \openone$. So
we obtain
\begin{equation}
  \label{eq:shamp}
  \mathcal{H}_{55}^{\vek{K}'} (\kk)
  = \mathcal{H}_{55}^{\vek{K}} (R_y^{-1} \kk) ,
\end{equation}
where $R_y^{-1} (k_x,k_y) = (-k_x,k_y)$.

SO coupling in the ideal material can be described by adding a
second-nearest-neighbor term \cite{and00, mar02, kan05a}
\begin{subequations}
  \label{eq:tb-hso}
  \begin{equation}
    \mathcal{H}_\mathrm{so} (\kvek) =
    \tvek[cc]{h_\mathrm{so} & 0 \\ 0 & - h_\mathrm{so}} ,
  \end{equation}
where
\begin{equation}
  \label{eq:tb-hso-fun}
  h_\mathrm{so} = i \lambda_\mathrm{so} \;
  \sum_{j,k}  \vek{s} \cdot (\vek{\tau}_1^{(j)} \times \vek{\tau}_1^{(k)})
  \:  e^{i\kvek\cdot(\vek{\tau}_1^{(j)} - \vek{\tau}_1^{(k)})} \,\, .
\end{equation}
\end{subequations}
Here $\vek{s}$ denotes the operator for spin angular
momentum. Similarly, the tight-binding model (\ref{eq:tb-ham}) can
be extended \cite{zar09} to include SO coupling due to the gradient
$e\vekc{E} \equiv \nabla V$ of an external potential $V$ by adding
a term
\begin{subequations}
  \label{eq:tb-rso}
  \begin{equation}
    \mathcal{H}_R (\kvek) =
    \tvek[cc]{0 & h_R \\ h_R^\ast & 0} \, ,
  \end{equation}
where
\begin{equation}
  \label{eq:tb-rso-fun}
  h_R = \lambda_R \; \vek{s} \cdot \vekc{E} \times
  \sum_j i\vek{\tau}_j \:  e^{i\kvek\cdot\vek{\tau}_j} \,\, .
\end{equation}
\end{subequations}
To lowest order in $\kk$ and $\vekc{E}$, the following additional terms
beyond Eq.\ (\ref{eq:sham}) are obtained:
\begin{equation}
  \label{eq:tb-so-k}
  \begin{array}[b]{rs{0.20em}>{\displaystyle}l}
   & \koeff{p}{55}{21} \, s_z \sigma_z
  + \koeff{r}{55}{61} \, s_z (\Ee_y \sigma_x - \Ee_x \sigma_y) \\[1ex] &
  + \koeff{r}{55}{62} \, \Ee_z (s_y \sigma_x - s_x \sigma_y) \\[1ex] &
  + \koeff{r}{55}{63} \, s_z [(k_x \Ee_y + k_y \Ee_x) \sigma_x
                      + (k_x \Ee_x - k_y \Ee_y) \sigma_y] \\[1ex] &
  + \koeff{r}{55}{64} \, \Ee_z [(s_x k_y + s_y k_x) \sigma_x
                        + (s_x k_x - s_y k_y) \sigma_y] ,
                   \end{array}
\end{equation}
where
\begin{subequations}
  \begin{eqnarray}
    \koeff{p}{55}{21} & = & \frack{3}{2} \lambda_{so} a^2 ,\\
    \koeff{r}{55}{61} & = & - \frack{\sqrt{3}}{2} \lambda_{R} a ,\\
    \koeff{r}{55}{62} & = & \frack{\sqrt{3}}{2} \lambda_{R} a ,\\
    \koeff{r}{55}{63} & = & \frack{1}{4} \lambda_{R} a^2 ,\\
    \koeff{r}{55}{64} & = & - \frack{1}{4} \lambda_{R} a^2 .
  \end{eqnarray}
\end{subequations}

While the above approach can be further extended in various ways, it
becomes difficult to explore all possible terms in a systematic
manner. For example, we will show below that the expansion
(\ref{eq:tb-so-k}) misses SO coupling terms of the form
\begin{equation}
  \label{eq:tb-so-kp}
   \koeff{r}{55}{11} \, (s_x k_y - s_y k_x) \Ee_z \openone
   + \koeff{r}{55}{12} \, s_z (k_x \Ee_y - k_y \Ee_x) \openone \, ,
\end{equation}
which illustrates the fact that qualitatively new terms may appear
if a more complete model beyond Eq.\ (\ref{eq:tb-rso}) is used to
describe Rashba spin-orbit coupling in graphene. From a different
perspective, we note that SO coupling, while small for systems made
of light atoms like carbon, can be expected to be underestimated by
the approach outlined above that expresses all properties in terms
of interatomic matrix elements. In general, the dominant effect of
the SO interaction is a coupling of the $p$ orbitals on the same
atom, \cite{cha77} which can be expected to contribute also to SO
coupling in graphene, see Sec.~\ref{sec:so}. (As usual, \cite{cas09}
the present approach has neglected the $p_x$ and $p_y$ orbitals.)

\section{Symmetry Analysis}
\label{sec:sym}

\subsection{Invariant Expansion}

While the power expansion (\ref{eq:sham}) of the tight-binding
Hamiltonian (\ref{eq:tb-hamp}) can be readily extended to arbitrary
orders of the wave vector $\kk$, it became noticeable in the above
discussion that it gets more difficult within this approach to
incorporate the effects of perturbations such as external electric
and magnetic fields, strain, or spin-orbit coupling. In contrast,
the theory of invariants \cite{bir74} enables systematic
construction of an invariant expansion for the effective Hamiltonian
$\mathcal{H} (\vekc{K})$ describing the electron states in the
vicinity of $\vek{K}$ and $\vek{K}'$ (or indeed any other point in
the Brillouin zone of any crystalline material). Here $\vekc{K}$
represents a general tensor operator, which can depend, e.g., on the
components of the kinetic wave vector $\kk$, on external electric
and magnetic fields $\vekc{E}$ and $\vek{B}$, on strain $\Strain$,
and on the intrinsic spin $\vek{s}$ of the electrons. Keeping in
mind that we want to extend the discussion from the spinless
Hamiltonian (\ref{eq:sham}) to a Hamiltonian that includes spin, we
first present a brief review of the general theory \cite{bir74}
before applying the formalism to our particular questions of
interest.

The Hamiltonian $\mathcal{H} (\vekc{K})$ is defined relative to a
set of basis functions $\Phi_\mu (\rr)$ transforming according to a
(reducible or irreducible) representation $\Gamma$ of the point
group $\mathcal{G}$. Denoting the transformation matrices by
$\mathcal{D}(g)$ ($g \in \mathcal{G}$), we have
\begin{equation}
  \label{eq:wf-trafo}
  \Phi_\mu (g^{-1} \, \rr)
  = \sum_{\mu'} \mathcal{D}_{\mu\mu'} (g) \, \Phi_{\mu'} (\rr) .
\end{equation}
Then the invariance of $\mathcal{H} (\vekc{K})$ under the symmetry
elements $g \in \mathcal{G}$ implies
\begin{equation}
  \label{eq:spatial-invar}
  \mathcal{D}(g) \, \mathcal{H} (g^{-1} \vekc{K}) \, \mathcal{D}^{-1}(g)
  = \mathcal{H} (\vekc{K}) .
\end{equation}
As shown in Ref.\ \onlinecite{bir74}, Eq.\ (\ref{eq:spatial-invar})
can be used to construct $\mathcal{H} (\vekc{K})$. This is greatly
simplified if we choose basis functions $\Phi_\mu (\rr)$
transforming according to \emph{irreducible} representations (IRs)
of $\mathcal{G}$. Then $\mathcal{H} (\vekc{K})$ can be decomposed
into blocks $\mathcal{H}_{\alpha\beta} (\vekc{K})$, where $\alpha$
and $\beta$ denote the spaces of the $n_\alpha$- and $n_\beta$-fold
degenerate basis functions, which transform according to the IRs
$\Gamma_\alpha$ and $\Gamma_\beta$ of $\mathcal{G}$. For each block
$\mathcal{H}_{\alpha\beta} (\vekc{K})$, one can find a complete set
of linearly independent $n_\alpha \times n_\beta$-dimensional
matrices ${X}_l^{(\kappa)}$ that transform according to the IRs
$\Gamma_\kappa$ (of dimension $L_\kappa$) contained in the product
representation $\Gamma_\alpha \times \Gamma_\beta^\ast$. Likewise,
$\vekc{K}$ can be decomposed into irreducible tensor operators
$\vekc{K}^{(\kappa',\lambda)}$ that transform according to the IRs
$\Gamma_{\kappa'}$ of $\mathcal{G}$ (Ref.~\onlinecite{subset}). Then
each block $\mathcal{H}_{\alpha\beta} (\vekc{K})$ can be written as
\begin{equation}
  \label{eq:invar}
  \mathcal{H}_{\alpha\beta} (\vekc{K}) 
  =  \sum_{\kappa, \, \lambda}
  \koeff{a}{\alpha\beta}{\kappa\lambda}
  \sum_{l=1}^{L_\kappa} {X}_l^{(\kappa)}
  \mathcal{K}_l^{(\kappa,\lambda) \, \ast} \, ,
\end{equation}
with material-specific coefficients
$\koeff{a}{\alpha\beta}{\kappa\lambda}$. By construction, each block
$\mathcal{H}_{\alpha\beta} (\vekc{K})$ is invariant under the
symmetry operations in $\mathcal{G}$ in the sense of
Eq.~(\ref{eq:spatial-invar}).

To proceed, we need to identify the symmetry of the eigenfunctions
(\ref{eq:K-wf}) at $\vek{K}$ and $\vek{K}'$ that form the basis
functions for $\mathcal{H} (\vekc{K}) $. The group of the wave
vector $\vek{K}$ is isomorphic to the trigonal point group
$\mathcal{G} = D_{3h}$ (while the point group of the honeycomb
structure is $D_{6h}$). Projection of $\Psi_{\vek{K}\lambda}$ on the
IRs of $D_{3h}$ shows \cite{bir74, slo55} that these functions transform
according to the two-dimensional IR $\Gamma_5$. \cite{kos63, koster_56}
More specifically, under the symmetry operations of $D_{3h}$, the Bloch
function $\Psi_{\vek{K},A (B)} (\rr)$ transforms like $\ket{\rho_{-
  (+)}}$, where $\ket{\rho_\mp} \equiv \frac{1}{\sqrt{2}}
\ket{\tilde{x} \mp i\tilde{y}}$. Here $\ket{\tilde{x}}$ and
$\ket{\tilde{y}}$ transform like the coordinate functions $x$ and
$y$ except that they do not change sign under inversion. Thus, from
a symmetry point of view, we may identify $\{\Psi_{\vek{K}\lambda}
(\rr)\}$ with $\{\ket{\rho_\mp}\}$. Note that
$\Psi_{\vek{K}'\lambda} = \Psi_{\vek{K}\lambda}^\ast$ [i.e., time
reversal does not cause extra degeneracies, see Eq.\
(\ref{eq:space-time}) below]. The role of $\ket{\rho_\mp}$ is thus
reversed at the $\vek{K}'$ point: $\ket{\rho_-}$ ($\ket{\rho_+}$)
corresponds to sublattice $B$ ($A$).

The TB model (\ref{eq:tb-hamp}) neglects the spin degree of freedom.
Thus we have obtained an ordinary IR ($\Gamma_5$ of $D_{3h}$), which
differs qualitatively from the double-group (spinor) IRs
characterizing a particle with a {\em genuine\/} spin-1/2 degree of
freedom coupled to a particle's orbital motion. The double group
$\mathcal{G}_d$ corresponding to a group $\mathcal{G}$ can be
written as $\mathcal{G}_d = \mathcal{G} \oplus \bar{E} \mathcal{G}$,
where $\bar{E}$ is a rotation by $2\pi$ around an arbitrary axis.
The presence of $\bar{E}$ in $\mathcal{G}_d$ reflects the well-known
fact \cite{sak85} that spin-1/2 spinors (which may be used
\cite{kos63} as basis functions for the spinor IR $\Gamma_7$ of
$D_{3h}$) change sign when rotated by $2\pi$. The basis functions of
$\Gamma_5$ do not change sign when rotated by $2\pi$, i.e.,
$\bar{E}$ acts like the neutral element $E$. In principle, this
could be studied experimentally in a setup similar to the neutron
interference experiments in Refs.~\onlinecite{rau75, wer75}.
However, such an experiment would obviously be complicated by the
fact that the electrons in graphene also carry a real spin degree of
freedom that is neglected in the present discussion. (See, however,
Sec.~\ref{sec:so} below.)

Choosing basis functions for $\Gamma_5$ that transform like
$\{\ket{\rho_-}, \ket{\rho_+}\}$, we obtain the basis matrices
listed in Table~\ref{tab:basemat}. Similarly, $\vekc{K}$ can be
decomposed into irreducible tensor operators
$\vekc{K}^{(\kappa,\lambda)}$ that transform according to the IRs
$\Gamma_\kappa$ of $D_{3h}$. Using the coordinate system in
Fig.~\ref{fig:lattice}(b), we get the lowest-order tensor operators
$\vekc{K}^{(\kappa,\lambda)}$ in Table~\ref{tab:tensorop}. We note
that given two irreducible tensor operators
$\vekc{K}^{(\alpha,\lambda_1)}$ and $\vekc{K}^{(\beta,\lambda_2)}$
transforming according to $\Gamma_\alpha$ and $\Gamma_\beta$, we get
new higher-order tensor operators transforming according to the IRs
$\Gamma_\kappa$ using the relation
\begin{equation}
  \label{eq:high-tensor}
  \mathcal{K}^{(\kappa,\lambda)}_l = \sum_{l_1, l_2}
  C^{\alpha\beta,\kappa}_{l_1 l_2, l} \;
  \mathcal{K}^{(\alpha,\lambda_1)}_{l_1}
  \mathcal{K}^{(\beta,\lambda_2)}_{l_2},
\end{equation}
where $C^{\alpha\beta,\kappa}_{l_1 l_2, l}$ denote the coupling
coefficients for $D_{3h}$. \cite{koster_56} We emphasize that this
approach guarantees that one obtains all irreducible tensor
operators up to a desired order. However, the definition of these
tensor operators is not unique because for two tensor operators
$\vekc{K}^{(\kappa,\lambda_1)}$ and $\vekc{K}^{(\kappa,\lambda_2)}$
which both transform according to $\Gamma_\kappa$ their
linear combination likewise transforms according to $\Gamma_\kappa$.

Combining the basis matrices and tensor operators according to Eq.\
(\ref{eq:invar}), we exactly reproduce Eq.\ (\ref{eq:sham}).
Third-order terms can be read off from Tables~\ref{tab:basemat}
and~\ref{tab:tensorop}. Terms of yet higher orders can be
constructed using Eq.\ (\ref{eq:high-tensor}). For $\vek{K}'$, we
obtain in generalization of Eq.\ (\ref{eq:shamp})
\begin{equation}
  \mathcal{H}_{55}^{\vek{K}'} (\vekc{K})
  = \mathcal{H}_{55}^{\vek{K}} (R_y^{-1}\vekc{K}) .
\end{equation}


\begin{table}
  \caption[]{\label{tab:basemat} Symmetrized matrices for the invariant
  expansion of the blocks $\mathcal{H}_{\alpha\beta}$ for the point group
  $D_{3h}$.}
$\arraycolsep 0.8em
\renewcommand{\arraystretch}{1.2}
\begin{array}{clcl} \hline \hline
\mbox{Block} &
\multicolumn{1}{l}{\mbox{Representations}} &
\multicolumn{2}{c}{\mbox{Symmetrized matrices}} \\ \hline

\mathcal{H}_{55} & \Gamma_5 \times \Gamma_5^\ast &
    \Gamma_1: & \openone  \\
& = \Gamma_1 + \Gamma_2 + \Gamma_6
  & \Gamma_2: & \sigma_z \\
  & & \Gamma_6: & \sigma_x, \sigma_y \\[1ex]

\mathcal{H}_{77} & \Gamma_7 \times \Gamma_7^\ast &
    \Gamma_1: & \openone \\
& = \Gamma_1 + \Gamma_2 + \Gamma_5
  & \Gamma_2: & \sigma_z \\
  & & \Gamma_5: & \sigma_x, -\sigma_y \\[1ex]

\mathcal{H}_{99} & \Gamma_9 \times \Gamma_9^\ast &
    \Gamma_1: & \openone \\
& = \Gamma_1 + \Gamma_2 + \Gamma_3 + \Gamma_4
  & \Gamma_2: & \sigma_z \\
  & & \Gamma_3: & \sigma_x \\
  & & \Gamma_4: & \sigma_y \\[1ex]

\mathcal{H}_{79} & \Gamma_7 \times \Gamma_9^\ast &
    \Gamma_5: & \openone, - i\sigma_z  \\
& = \Gamma_5 + \Gamma_6
  & \Gamma_6: & \sigma_x, \sigma_y \\ \hline \hline
\end{array}$
\end{table}

\begin{table}
  \caption[]{\label{tab:tensorop}
   Irreducible tensor components for the point group $D_{3h}$. Those
   printed in bold give rise to invariants in
   $\mathcal{H}_{55}^\vek{K} (\vekc{K})$ allowed by time-reversal invariance.
   (Terms proportional to $k_z$ or $\strain_{jz}$ are not listed as they are
   irrelevant for graphene.)\\
   Notation: $\{A, B\} \equiv \frac{1}{2} (AB + BA)$.}
  \begin{textmath}
  \extrarowheight 0.2ex
  \begin{array}{cs{1.5em}l} \hline \hline
    \Gamma_1 & \allowed{1}; \;
    \allowed{k_x^2 + k_y^2}; \;
    \allowed{\{k_x, 3k_y^2 - k_x^2\}}; \;
    \forbidden{k_x \Ee_x + k_y \Ee_y};  \\ &
    \allowed{\strain_{xx} + \strain_{yy}}; \;
    \allowed{(\strain_{yy} - \strain_{xx}) k_x + 2 \strain_{xy} k_y}; \\ &
    \forbidden{(\strain_{yy} - \strain_{xx}) \Ee_x + 2 \strain_{xy} \Ee_y}; 
    \allowed{s_x B_x + s_y B_y}; \;
    \allowed{s_z B_z}; \\ &
    \allowed{(s_x k_y - s_y k_x) \Ee_z}; \;
    \allowed{s_z (k_x \Ee_y - k_y \Ee_x)}; \\
    \Gamma_2 & \forbidden{\{k_y, 3k_x^2-k_y^2\}}; \;
    \allowed{B_z}; \;
    \allowed{k_x \Ee_y - k_y \Ee_x}; \\ &
    \forbidden{(\strain_{xx} - \strain_{yy}) k_y + 2 \strain_{xy} k_y}; \;
    \allowed{(\strain_{xx} + \strain_{yy}) B_z}; \\ &
    \allowed{(\strain_{xx} - \strain_{yy}) \Ee_y + 2 \strain_{xy} \Ee_x}; \;
    \allowed{s_z}; \;
    \forbidden{s_x B_y - s_y B_x}; \\ &
    \forbidden{(s_x k_x + s_y k_y) \Ee_z}; \;
    \allowed{s_z (\strain_{xx} + \strain_{yy})}; \\
    \Gamma_3 & B_x k_x + B_y k_y; \; \Ee_x B_x + \Ee_y B_y; \; \Ee_z B_z; \\ &
    (\strain_{yy} - \strain_{xx}) B_x + 2 \strain_{xy} B_y; \;
    s_x k_x + s_y k_y; \\ & s_x \Ee_x + s_y \Ee_y; \; s_z \Ee_z; \;
    s_x (\strain_{yy} - \strain_{xx}) + 2 s_y \strain_{xy} \\
    \Gamma_4 & B_x k_y - B_y k_x; \; \Ee_z; \;
    \Ee_x B_y - \Ee_y B_x; \\ &
    (\strain_{xx} - \strain_{yy}) B_y + 2 \strain_{xy} B_x; \;
    (\strain_{xx} + \strain_{yy}) \Ee_z; \\ &
    s_x k_y - s_y k_x; \; s_x \Ee_y - s_y \Ee_x; \;
    s_y (\strain_{xx} - \strain_{yy}) + 2 s_x \strain_{xy} \\
    \Gamma_5 & B_x, B_y; \; B_y k_y - B_x k_x, B_x k_y + B_y k_x;  \;
    k_y \Ee_z, - k_x \Ee_z; \\ &
    \Ee_y B_y - \Ee_x B_x, \Ee_y B_x + \Ee_x B_y; \; 
    (\strain_{xx} + \strain_{yy}) (B_x, B_y); \\ &
    (\strain_{xx} - \strain_{yy}) B_x + 2 \strain_{xy} B_y,
    (\strain_{yy} - \strain_{xx}) B_y + 2 \strain_{xy} B_x; \\ &
    2 \strain_{xy} \Ee_z, (\strain_{xx} - \strain_{yy}) \Ee_z; \;
    s_x, s_y; \\ & s_y k_y - s_x k_x, s_x k_y + s_y k_x; \;
    s_y B_z, - s_x B_z; \\ & s_z B_y, - s_z B_x; \;
    s_y \Ee_y - s_x \Ee_x, s_x \Ee_y + s_y \Ee_x; \\ &
    (s_x, s_y) (\strain_{xx} + \strain_{yy}); \\ &
    s_x (\strain_{xx} - \strain_{yy}) - 2 s_y \strain_{xy},
    s_y (\strain_{yy} - \strain_{xx}) - 2 s_x \strain_{xy} \\
    \Gamma_6 & \allowed{k_x, k_y}; \;
    \allowed{\{k_y+k_x, k_y - k_x\}, 2 \{k_x, k_y\}}; \\ & 
    \allowed{\{k_x, k_x^2 + k_y^2\}, \{k_y, k_x^2 + k_y^2\}}; \;
    \forbidden{B_z k_y, - B_z k_x}; \\ &
    \forbidden{\Ee_x, \Ee_y}; \;
    \forbidden{k_y \Ee_y - k_x \Ee_x, k_x \Ee_y + k_y \Ee_x}; \\ &
    \allowed{\Ee_y B_z, - \Ee_x B_z}; \;
    \allowed{\Ee_z B_y, - \Ee_z B_x}; \\ &
    \allowed{\strain_{yy} - \strain_{xx}, 2 \strain_{xy}}; \;
    \allowed{(\strain_{xx} + \strain_{yy}) (k_x, k_y)}; \\ &
    \allowed{(\strain_{xx} - \strain_{yy}) k_x + 2\strain_{xy} k_y,
             (\strain_{yy} - \strain_{xx}) k_y + 2\strain_{xy} k_x}; \\ &
    \forbidden{2 \strain_{xy} B_z, (\strain_{xx} - \strain_{yy}) B_z}; \\ &
    \forbidden{(\strain_{xx} - \strain_{yy}) \Ee_x + \strain_{xy} \Ee_y,
               (\strain_{yy} - \strain_{xx}) \Ee_y + \strain_{xy} \Ee_x}; \\ &
    \forbidden{(\strain_{xx} + \strain_{yy}) (\Ee_x, \Ee_y)}; \;
    \forbidden{s_z k_y, - s_z k_x}; \\ &
    \allowed{s_y B_y - s_x B_x, s_x B_y + s_y B_x}; \;
    \allowed{s_z \Ee_y, - s_z \Ee_x}; \\ &
    \allowed{s_y \Ee_z, - s_x \Ee_z}; \;
    \allowed{s_z (k_x \Ee_y + k_y \Ee_x), s_z (k_x \Ee_x - k_y \Ee_y)}; \\ &
    \allowed{(s_x k_y + s_y k_x) \Ee_z, (s_x k_x - s_y k_y) \Ee_z}; \\ &
    \forbidden{2 s_z \strain_{xy}, s_z (\strain_{xx} - \strain_{yy})}; \;
    \\ \hline \hline
  \end{array}
  \end{textmath}
\end{table}

The part of $\mathcal{H}_{55}^\vek{K}$ linear in $\kk$ is formally
equivalent to the Dirac Hamiltonian for massless (chiral) fermions.
\cite{sak67} Also, it is related via a simple unitary transformation
with the Dresselhaus \cite{dre55a} and the Rashba \cite{byc84} terms
in quasi-2D systems. \cite{win03} All these models give rise to a
dispersion that is linear in the limit of small $\kk$. Yet for each
of these models, the transformational properties of the basis
functions under the corresponding symmetry operations are
qualitatively different. \cite{envelope} The Dirac equation reflects
the continuous symmetries of the Lorentz group. \cite{dirac} In
quasi-2D systems, the $k$-linear Dresselhaus term is intimately
related with the tetrahedral symmetry of the zinc blende structure,
\cite{dre55a} whereas the Rashba term emerges from a model with
axial \cite{win03} (or hexagonal \cite{ras60}) symmetry. These terms
refer to electron states transforming according to spinor
representations of the corresponding crystallographic point groups,
and they are nonzero only as a consequence of spin-orbit
coupling. \cite{win03} In contrast, the Hamiltonian
$\mathcal{H}_{55}^\vek{K}$ refers to the basis functions (\ref{eq:K-wf}).
It is applicable to spinless particles or particles for which the
spin degree of freedom is decoupled from the orbital motion, and the
group element $\bar{E}$ does \emph{not} play a role. \cite{wurtzite}
From a more general perspective, we see here that the symmetries of
a system determine the invariant expansion (\ref{eq:invar}) and the
band structure $E_n (\kk)$. Yet it is, in general, not possible to
follow the opposite path and infer the symmetries of a system from
the Hamiltonian and the band structure $E_n (\kk)$.

\subsection{Implications of Time Reversal Invariance}

Time-reversal invariance results in additional constraints for the
allowed terms in the invariant expansion (\ref{eq:invar}). Time
reversal corresponds to complex conjugation. Focusing here on the
important case (usually \cite{bir74} denoted ``case a'') that we
have a linear relation between the complex-conjugate basis functions
$\Phi_\lambda^\ast$ and the original basis functions~$\Phi_\lambda$,
\begin{equation}
  \label{eq:wf-t-trafo}
  \Phi_\lambda^\ast (\rr)
  = \sum_{\lambda'} \mathcal{T}_{\lambda\lambda'} \Phi_{\lambda'} (\rr),
\end{equation}
time-reversal invariance implies
\begin{equation}
  \label{eq:time-invar}
  \mathcal{T}^{-1} \, \mathcal{H} (\zeta \vekc{K}) \, \mathcal{T}
  = \mathcal{H}^\ast (\vekc{K}) = \mathcal{H}^t (\vekc{K}) .
\end{equation}
Here, $\ast$ denotes complex conjugation and $t$ transposition. The
prefactor $\zeta$ depends on the behavior of $\vekc{K}$ under time
reversal. $\kk$, $\vek{B}$, and $\vek{s}$ are odd under time
reversal so that then $\zeta=-1$, while $\vekc{E}$ and $\Strain$
have $\zeta=+1$.

The general analysis needs to be modified due to the fact that the
valleys $\vek{K}$ and $\vek{K}'$ are inequivalent points in the star
$\{\kvek\}$ that characterizes the IRs of the space group for these
values of $\kvek$. Therefore, the eigenstates at $\vek{K}$ and
$\vek{K}'$ need to be combined in order to use the general relation
(\ref{eq:wf-t-trafo}). It turns out that the bands at $\vek{K}$ and
$\vek{K}'$ belong to the case denoted ``$a_2$'' in
Ref.~\onlinecite{bir74}, i.e., the functions $\theta \Psi_{\vek{K}
 \lambda} = \Psi_{\vek{K} \lambda}^\ast$ at $\vek{K}$ and
$\Psi_{\vek{K}' \lambda}$ at $\vek{K}'$ are linearly related via a
unitary matrix~$\mathcal{T}$
\begin{equation}
  \label{eq:time}
  \theta \, \Psi_{\vek{K},\lambda} = \Psi_{\vek{K}\lambda}^\ast
  = \sum_{\lambda'} \mathcal{T}_{\lambda\lambda'}
  \: \Psi_{\vek{K}'\lambda'} \,,
\end{equation}
where $\theta$ denotes the time-reversal operator. But the
crystallographic point group $D_{6h}$ also contains elements $R$
that likewise establish linear relations between $\Psi_{\vek{K}
 \lambda}$ and $\Psi_{\vek{K}' \lambda}$, see Eq.\
(\ref{eq:K-Kp-wf}). Combining Eqs.\ (\ref{eq:K-Kp-wf}) and
(\ref{eq:time}), we obtain the linear relation
\begin{equation}
  \label{eq:space-time}
  \theta \, \Psi_{\vek{K},\lambda} = \Psi_{\vek{K}\lambda}^\ast
  = \sum_{\lambda',\lambda''} \mathcal{T}_{\lambda\lambda'} \:
  \mathcal{D}_{\lambda'\lambda''}(R) \: \Psi_{\vek{K}\lambda''} \,.
\end{equation}
Obviously, the matrix $\mathcal{T}$ depends on the symmetry element
$R$. For $R=R_y$ (which preserves the sublattices $A,B$ as discussed
above) we get $\mathcal{T} = \openone$. If we choose $R$ as
inversion [which flips the sublattices $A,B$, thus $\mathcal{D}(R) =
\sigma_x$] we get $\mathcal{T} = \sigma_x$ (as in
Ref.~\onlinecite{man07}). However, in any case we have
$\mathcal{T}^2=+1$, as expected for spinless particles. \cite{sak85}

Combining the general relation (\ref{eq:time-invar}) with
(\ref{eq:space-time}) we obtain the condition \cite{bir74,man07}
\begin{equation}
  \label{eq:timea2}
  \mathcal{T}^{-1} \mathcal{H} (R^{-1} \vekc{K}) \mathcal{T}
  = \mathcal{H}^\ast (\zeta \vekc{K}) = \mathcal{H}^t (\zeta \vekc{K}) .
\end{equation}
It provides a general criterion for determining which terms in the
expansion (\ref{eq:invar}) are allowed by time-reversal invariance
and which terms are forbidden.

\section{Discussion of Invariant Expansion}
\label{sec:disc}

Tables~\ref{tab:basemat} and~\ref{tab:tensorop} are the main result
of this work. In the following, we discuss some associated physical
consequences. In Sec.~\ref{sec:orb}, we focus on the orbital motion
of the electrons, including the effect of perturbing electric and
magnetic fields $\vekc{E}$ and $\vek{B}$. In Sec.~\ref{sec:strain},
we consider the effect of strain. Finally, the effect of spin-orbit
coupling is discussed in Sec.~\ref{sec:so}.

We use different letters for the prefactors in the invariant
expansion (\ref{eq:invar}) to classify the terms according to their
relevance, though such a scheme cannot be rigorous when mixed
effects are considered. Coefficients
$\koeff{a}{\alpha\beta}{\kappa\lambda}$ refer to invariants
characterizing the orbital motion in the absence of fields;
$\koeff{r}{\alpha\beta}{\kappa\lambda}$ denote the prefactors of
orbital or spin-dependent Rashba-like invariants proportional to an
external electric field; $\koeff{z}{\alpha\beta}{\kappa\lambda}$
denote prefactors of Zeeman-like terms proportional to $B$;
coefficients $\koeff{b}{\alpha\beta}{\kappa\lambda}$ characterize
the Bir-Pikus strain Hamiltonian for graphene; and coefficients
$\koeff{p}{\alpha\beta}{\kappa\lambda}$ characterize the intrinsic
(Pauli) spin-orbit coupling.

\subsection{Orbital Motion}
\label{sec:orb}

Equation (\ref{eq:timea2}) implies that all terms in Eq.\
(\ref{eq:sham}) are allowed by time-reversal invariance, yet, e.g.,
the third-order invariant $k_y (3k_x^2-k_y^2) \, \sigma_z$ is
forbidden. The tensor operators that give rise to invariants in
$\mathcal{H}_{55}^\vek{K} (\vekc{K})$ allowed by time reversal have
been printed in bold in Table~\ref{tab:tensorop}.

The existence of $k$-linear terms in $\mathcal{H}_{55}^\vek{K}$ is
intimately related to the behavior of $\mathcal{H}_{55}^\vek{K}$
under time reversal characterized by Eq.\ (\ref{eq:timea2}).
\cite{bir74, man07} For comparison, consider the $\Gamma$ point $\kk
= 0$ of a material with point group $D_{3h}$, where
Tables~\ref{tab:basemat} and~\ref{tab:tensorop} are valid, too. Yet
the $k$-linear terms in $\mathcal{H}_{55}^\vek{K}$ are then
forbidden because, in this case (case $a_1$ in
Ref.~\onlinecite{bir74}), the matrices $\{{X}_l^{(\kappa)}\}$ can be
classified as even or odd under time reversal with $\{{X}_l^{(6)}\}$
even.

To linear order of $\Ee_z$ and in the absence of other
perturbations, an external perpendicular electric field $\Ee_z$
cannot couple to the planar orbital motion in graphene (to all
orders in $k$). The lowest-order invariant involving in-plane
electric fields reads
\begin{equation}\label{eq:elecField}
  \koeff{r}{55}{21} (k_x \Ee_y - k_y \Ee_x ) \sigma_z \quad .
\end{equation}
A linear Zeeman splitting in a parallel magnetic field
$\vek{B}_\|= (B_x,B_y,0)$ (and independent of $k$) is likewise
forbidden by symmetry. However, this result is not a consequence of
the planar geometry of graphene. Indeed, it follows from the
character tables of $D_{3h}$ that an in-plane orbital magnetic
moment is absent in all systems characterized by this point group,
e.g., also in graphite. On the other hand, the allowed
invariant~\cite{man07}
\begin{equation}
  \koeff{z}{55}{21} B_z \sigma_z
\end{equation}
implies that electrons in graphene have an orbital magnetic moment
in $z$ direction. We have here an interesting difference between the
Dirac-like orbital motion of Bloch electrons in graphene and truly
relativistic systems. For neutrinos, which are almost-massless Dirac
fermions, the magnetic moment is proportional to their mass.
\cite{zub03} The spin magnetic moment of Rashba and 2D Dresselhaus
electrons has both a $z$ and an in-plane component, though generally
these are different. \cite{win03}

In a field $B_z > 0$ we can utilize the usual \cite{win03} ladder
operators $a^\pm = l_B (k_x \pm i k_y)/\sqrt{2}$ with magnetic
length $l_B=\sqrt{\hbar/|e B_z|}$ to obtain up to linear order of
$\kk$ and $B_z$ [$\sigma_\pm \equiv \frac{1}{2} (\sigma_x \pm
i\sigma_y)$]
\begin{equation}
  \mathcal{H}_{55}^\mathrm{L\pm} = \koeff{a}{55}{10} \openone
  + \frac{\sqrt{2}\, \koeff{a}{55}{61}}{l_{\text{B}}}
  \left( a^+ \sigma_\mp + a^- \sigma_\pm \right)
   \pm \koeff{z}{55}{21} \, B_z \, \sigma_z \, ,
\end{equation}
where the upper (lower) sign applies to $\vek{K}$ ($\vek{K}'$). Both
$\mathcal{H}_{55}^\mathrm{L+}$ and $\mathcal{H}_{55}^\mathrm{L-}$
have the same Landau spectrum
\begin{subequations}
  \label{eq:newLL}
  \begin{eqnarray}
    E_{n\pm} &=& \koeff{a}{55}{10}
    \pm \frac{\sqrt{2} \, \koeff{a}{55}{61}}{l_{\text{B}}}
    \,\sqrt{n +\left( \frac{\koeff{z}{55}{21}}{\koeff{a}{55}{61}}\right)^2 
    \left| \frac{\hbar B_z}{2 e} \right| } \, , \\
    E_0 &=& \koeff{a}{55}{10} - \koeff{z}{55}{21}\, \left| B_z \right| \, ,
  \end{eqnarray}
\end{subequations}
with positive integers $n$. Finding $E_0\ne \koeff{a}{55}{10}$ signals
broken particle-hole symmetry. For $\koeff{z}{55}{21}=0$, Eq.\
(\ref{eq:newLL}) is identical to the spectrum of 2D massless Dirac
fermions in a magnetic field $B_z$ (Ref.~\onlinecite{mcc56}). The
spectrum (\ref{eq:newLL}) is obtained also for the Rashba
\cite{byc84} and 2D Dresselhaus \cite{win03} models in the limit of
an infinite effective electron mass. This result illustrates the
fact that the spectrum (\ref{eq:newLL}) for $B_z > 0$ and
$\koeff{z}{55}{21}=0$ is determined by the dispersion (\ref{eq:disp}) at
$B=0$ but does not depend on the transformational properties of the
underlying basis functions (which are different for these models).

\subsection{Strain-induced effects}
\label{sec:strain}

To incorporate the effect of strain $\Strain$ into the invariant
expansion (\ref{eq:invar}), we follow the general theory developed in
Ref.~\onlinecite{bir74}. A small homogeneous strain for a 2D sheet
such as graphene is defined by the symmetric strain tensor
\cite{lan7e}
\begin{equation}
  \label{eq:strain-def}
  \strain_{ij} = \frac{1}{2} \left(
    \frac{\partial u_i}{\partial r_j}
    + \frac{\partial u_j}{\partial r_i}
    + \frac{\partial u_z}{\partial r_i}\, \frac{\partial u_z}{\partial r_j}
  \right) ,
\end{equation}
where $\vek{u} (\vek{r})$ is the displacement vector at point
$\vek{r}$ due to strain and $i,j\in\{x,y\}$. Note that, while the position
vector $\vek{r}$ is two-dimensional, the displacement $\vek{u} (\vek{r})$
can have three nonzero components. Nevertheless, to lowest order, the
components $\strain_{jz}$ and $\strain_{zz}$ of the general
(three-dimensional) strain tensor vanish in a 2D sheet like
graphene; \cite{lan7e} only the in-plane components of $\Strain$ are finite.
Recent experimental studies have mapped \cite{tea09} and engineered
\cite{bao09} strain in single-layer graphene.

Quite generally, \cite{bir74} the components $\strain_{ij}$ of the
strain tensor transform like the symmetrized products $\{k_i,k_j\}$.
Thus to lowest order, we get the irreducible tensor components
proportional to $\strain_{ij}$ listed in Table~\ref{tab:tensorop}.
Previous work~\cite{suz02a, man07a} has already identified the terms
\begin{equation}
 \koeff{b}{55}{11} (\strain_{xx} + \strain_{yy})\openone
 + \koeff{b}{55}{61} [ (\strain_{yy} - \strain_{xx})
   \sigma_x + 2 \strain_{xy} \sigma_y],
\end{equation}
where the second term can be interpreted as
arising from a geometry-related fictitious vector potential.
\cite{kan97,voz10} From that viewpoint, the term
\begin{equation}
  \label{eq:quadstrain}
  \koeff{b}{55}{12}
  [(\strain_{yy} - \strain_{xx}) k_x + 2 \strain_{xy} k_y] \openone
\end{equation}
has a straightforward interpretation in terms of the same type of
gauge-field correction to the quadratic-dispersion contribution
$\koeff{a}{55}{11} (k_x^2 + k_y^2)\openone$. Possibilities to use
strain-induced pseudo-magnetic fields to manipulate electronic
transport in graphene have attracted significant attention recently.
\cite{fog08, per09, gui10} Furthermore, an isotropic renormalization
of the electron velocity as embodied in the term
\begin{equation}
  \koeff{b}{55}{62}
  (\strain_{xx} + \strain_{yy})(k_x \sigma_x + k_y \sigma_y) 
\end{equation}
was discussed in conjunction with smooth rippling of the graphene
sheet. \cite{jua07} Our results suggest that even an anisotropic
velocity renormalization can be engineered using strain, based on
the contribution
\begin{equation}\label{eq:newstrain1}
  \koeff{b}{55}{63}\{ [(\strain_{xx} - \strain_{yy}) k_x + 2\strain_{xy} k_y]
  \sigma_x + [(\strain_{yy} - \strain_{xx}) k_y + 2\strain_{xy} k_x]
  \sigma_y \} .
\end{equation}
This mechanism for creating an anisotropic dispersion provides an
alternative to the previously suggested~\cite{par08} periodic
modulation of graphene sheets.

Several terms involve strain in combination with external fields. A
strain-dependent renormalization of orbital Zeeman coupling is given
by
\begin{equation}\label{eq:newstrain2}
  \koeff{b}{55}{21} (\strain_{xx} + \strain_{yy}) B_z \sigma_z  \quad .
\end{equation}
In-plane electric fields coupled with strain generate a gap via the
contribution
\begin{equation}\label{eq:newstrain3}
  \koeff{b}{55}{22} [ (\strain_{xx} - \strain_{yy}) \Ee_y + 2 \strain_{xy}
  \Ee_x ] \sigma_z \quad .
\end{equation}
Both the orbital $g$-factor renormalization and the gap size could
vary randomly in space, as strain is associated with certain types
of disorder such as ripples. \cite{gui08}

Our symmetry analysis has yielded all the terms that can be
generated from Eqs.~(\ref{eq:sham}) and (\ref{eq:elecField}) by
replacing
\begin{equation}
  \label{eq:strain-gauge}
  k_x \to k_x + \alpha (\epsilon_{yy} - \epsilon_{xx}), \qquad
  k_y \to k_y + 2 \alpha \,\epsilon_{xy} ,
\end{equation}
thus resembling a minimal coupling to a strain-related geometric
gauge field with coupling constant $\alpha$. This result reflects
the fact that both the wave vector components $k_x, k_y$ and the
strain tensor components $\strain_{xx} - \strain_{yy}, 2
\strain_{xy}$ transform according to $\Gamma_6$ (see
Table~\ref{tab:tensorop}) so that these terms may be combined as in
Eq.\ (\ref{eq:strain-gauge}) to form a new (alternative) tensor
operator. However, the construction of higher-order tensor operators
using Eq.\ (\ref{eq:high-tensor}) requires that we include not only
the weighted sum of $k_x, k_y$ and $\strain_{xx} - \strain_{yy}, 2
\strain_{xy}$ as one tensor operator, as in Eq.\
(\ref{eq:strain-gauge}). Rather, their (weighted) difference also
constitutes a linearly independent tensor operator. So even if the
latter has been defined such that the prefactor of the corresponding
linear-order invariant vanishes, it is not guaranteed that the
prefactors of higher-order invariants constructed via Eq.\
(\ref{eq:high-tensor}) will vanish as well. Thus the simple
replacement (\ref{eq:strain-gauge}) may be insufficient to account
for strain effects in higher orders.

\subsection{Spin-Orbit Coupling}
\label{sec:so}

As discussed in Ref.~\onlinecite{bir74}, spin-orbit (SO) coupling
can be incorporated in the invariant expansion (\ref{eq:invar}) in
two equivalent ways. In the first approach, the components of the
pseudovector $\vek{s}$ enter the general tensor $\vekc{K}$ in much
the same way as $\kk$, $\vek{B}$, $\vekc{E}$, and $\Strain$. In
Table~\ref{tab:tensorop}, we have listed the resulting lowest-order
irreducible tensor operators. With spin taken into account in this
way, the basis functions of the Hamiltonian
$\mathcal{H}_{55}^\vek{K} (\vekc{K})$ transform according to the
direct product $\Gamma_5 \times \mathcal{D}_{1/2}$ of the
representation $\Gamma_5$ according to which the coordinate
functions (\ref{eq:K-wf}) transform and the representation
$\mathcal{D}_{1/2}$ of $SU(2)$ according to which the spin functions
$\ket{\uparrow}$ and $\ket{\downarrow}$ transform. In other words,
the basis functions are constructed from the eigenfunctions without
$\mathcal{H}_\mathrm{so}$, and $\mathcal{H}_\mathrm{so}$ is
introduced as a perturbation (similar to $\vekc{E}$, $\vek{B}$, and
$\Strain$). One can thus ascertain at once which of the coefficients
$\koeff{a}{55}{\kappa\lambda}$ entering $\mathcal{H}_{55}^\vek{K}
(\vekc{K})$ are relativistically small. In lowest order, we get the
terms given in Eqs.\ (\ref{eq:tb-so-k}) and (\ref{eq:tb-so-kp}), as
well as several invariants where strain couples to the intrinsic
spin degree of freedom. As noted above, terms shown in Eq.\
(\ref{eq:tb-so-kp}) were previously omitted.

Alternatively, we can construct $\mathcal{H} (\vekc{K})$ by directly
using the double-group representations of $D_{3h}$ contained in the
product representation $\Gamma_5 \times \mathcal{D}_{1/2}$.
Decomposing $\Gamma_5 \times \mathcal{D}_{1/2}$ into IRs gives rise
to spinors that transform according to the double-group IRs
$\Gamma_7$ (with representative basis functions $\{ \ket{\rho_-
 \uparrow}, \ket{\rho_+ \downarrow} \}$) and $\Gamma_9$ ($\{
\ket{\rho_- \downarrow}, \ket{\rho_+ \uparrow} \}$) of $D_{3h}$. The
corresponding basis matrices are also listed in
Table~\ref{tab:basemat} which are again combined with the tensor
operators in Table~\ref{tab:tensorop}. Thus we obtain the $4\times
4$ Hamiltonian
\begin{equation}
  \label{eq:dham}
  \mathcal{H} = \left( \begin{array}{cc}
   \mathcal{H}_{77} & \mathcal{H}_{79} \\
   \mathcal{H}_{97} & \mathcal{H}_{99}
    \end{array}\right) \, ,
\end{equation}
where each block $\mathcal{H}_{\alpha\beta}$ is given by an
invariant expansion of the form (\ref{eq:invar}). Obviously, both
approaches are related by a unitary transformation.

The most important consequence of SO coupling, which can be inferred
directly from Eq.\ (\ref{eq:dham}), is the opening of a gap $\Delta$
between the bands $\Gamma_7$ and $\Gamma_9$, so that we get
$\koeff{a}{77}{10} = \epsilon_{2p} + \frac{1}{2}\Delta$ and
$\koeff{a}{99}{10} = \epsilon_{2p} - \frac{1}{2}\Delta$
(Refs.~\onlinecite{slo58, dre65}). The origin of this gap can be
understood as follows. We may replace the spinless basis functions
$\ket{\rho_\mp}$ by the unitarily equivalent basis functions
$\ket{\tilde{x}} = \frack{1}{\sqrt{2}} (\ket{\rho_-} +
\ket{\rho_+})$ and $\ket{\tilde{y}} = \frack{i}{\sqrt{2}}
(\ket{\rho_-} - \ket{\rho_+})$. The latter basis functions have
equal magnitudes on both sublattices $A$ and $B$. Then we have
$\Delta = \frac{\hbar}{2m_0^2c^2} \bra{\tilde{x}} [\nabla V \times
\vek{p}]_z \ket{\tilde{y}}$, where $V$ is the microscopic crystal
potential of graphene. From a symmetry point of view, the gap
$\Delta$ is thus analogous to the gap that separates the topmost
valence band in semiconductors such as Ge and GaAs from the
spin-split-off valence band. \cite{win03} Of course, in graphene we
would have $\Delta = 0$ if the basis functions $\ket{\tilde{x}}$ and
$\ket{\tilde{y}}$ were comprised of pure $\pi$ ($p_z$) orbitals.
However, SO coupling induces a mixing of the $\pi$ ($p_z$) and
$\sigma$ ($p_x,p_y$) orbitals in graphene that contributes to
$\Delta$ in second order of SO coupling. \cite{slo55, min06, hue06}
A second contribution to $\Delta$ (linear in SO coupling) stems from
an SO-induced coupling between the atomic $p_z$ and higher atomic
orbitals such as $d$ states. \cite{slo55, slo58} These mechanisms
refer to SO matrix elements for states localized on the same atom.
As SO coupling originates from the steep gradients of the Coulomb
potentials in the atomic cores, these terms generally provide the
dominant effect, \cite{cha77} though we see that such mechanisms are
less effective in graphene. A third contribution to $\Delta$ is due
to the second-nearest-neighbor coupling of the $p_z$ orbitals
discussed in Ref.~\onlinecite{kan05a}. Recent first-principles
calculations \cite{gmi09} found $\Delta \approx 24~\mathrm{\mu eV}$,
with this gap arising almost entirely from contributions from $d$
and higher orbitals. The sign of $\Delta$ and, thus, the order of
the bands $\Gamma_7$ and $\Gamma_9$ cannot be inferred from our
analysis.

Unlike the case of inversion-asymmetric crystal structures such as
zinc blende and wurtzite, SO coupling in inversion-symmetric graphene
does not give rise to spin splitting. Hence, with spin taken into account,
we get a two-fold spin degeneracy for all bands throughout the Brillouin
zone. \cite{bir74} Accordingly, the Hamiltonian (\ref{eq:dham}) preserves
the two-fold spin degeneracy (for $B = \Ee = 0$). An external field
$\vekc{E}$ breaks the spatial inversion symmetry.
Tables~\ref{tab:basemat} and~\ref{tab:tensorop} show that, to lowest
order in $k$, the resulting spin splitting is due to the invariants given in
Eqs.\ (\ref{eq:tb-so-k}) and (\ref{eq:tb-so-kp}).

As is the case in other materials, \cite{bir74} strain mediates a
coupling between intrinsic-spin and orbital dynamics. The
lowest-order contribution
\begin{equation}\label{eq:so_strain}
  \koeff{p}{55}{22} (\strain_{xx} + \strain_{yy}) s_z \sigma_z
\end{equation}
constitutes a renormalization of the intrinsic SO coupling
$\koeff{p}{55}{21} s_z \sigma_z$. The existence of this term enables
strain engineering of spin splitting in graphene. It also implies
that disorder associated with strain (such as ripples) gives rise to
a spatially random SO coupling that should have implications
for spin relaxation in graphene beyond previously considered
mechanisms. \cite{hue09}

\section{Conclusions}
\label{sec:conc}

We have performed a detailed symmetry analysis of the trigonal band
structure of graphene. A systematic invariant expansion of the
envelope-function Hamiltonians describing electron states near the
$\vek{K}$ and $\vek{K}'$ points is presented, including effects due
to external electric and magnetic fields, strain, and spin-orbit
coupling. Our results include all terms up to second order as well
as selected terms up to third order and they include several
previously unnoticed terms. Examples for the latter are shown in
Eqs.~(\ref{eq:tb-so-kp}), (\ref{eq:elecField}),
(\ref{eq:newstrain1})--(\ref{eq:newstrain3}), and
(\ref{eq:so_strain}). We have also highlighted the peculiar role
played by time-reversal invariance in determining graphene's
band structure.

It should be noted that, in principle, our analysis based on the
invariant expansion (\ref{eq:invar}) could be extended to multilayer
graphene. However, as discussed, e.g., in Ref.~\onlinecite{mal09},
$N$-layer graphene systems consisting of an even (odd, $N>1$) number
of sheets have the point group $D_{3d}$ ($D_{3h}$), and the
$\vek{K}$ points have the point group $D_3$ ($C_{3h}$). The proper
symmetry analysis for multilayer graphene needs to be based on these
symmetries and, hence, will differ qualitatively from the one
presented in this work.

Before closing, we comment on the significance of the different
transformational properties characterizing the wave functions of
electrons in graphene and massless Dirac fermions, respectively. Our
group-theoretical analysis suggests to divide the electronic
properties of these systems into two categories: (I)~those that
emerge from the linear dispersion (\ref{eq:disp}) but that are
independent of the transformational properties of the basis
functions, and (II)~those that \emph{do\/} reflect these
transformational properties. \cite{envelope} Clearly, the
experimentally verified \cite{nov05a, zha05c} Landau spectrum
(\ref{eq:newLL}) belongs to category (I). Similarly,
\emph{Zitterbewegung}-like effects, \cite{win07} i.e., phenomena
arising due to the interference between electron states from
neighboring bands, \cite{cul08a} generally belong to category (I).
In contrast, the magnetic moment belongs to category (II).
Similarly, Kramers' degeneracy only holds for particle states
transforming according to spinor IRs. In this context, it is not
important whether the symmetry group is discrete or continuous, as
in both cases we can distinguish ordinary and spinor IRs.


\begin{acknowledgments}
  The authors appreciate stimulating discussions with C.-S.~Chu,
  D.~Culcer, E.~I.\ Rashba, A.~I.\ Signal, and L.-Y.\ Wang. This
  work is supported by the Marsden Fund Council (contract MAU0702)
  from New-Zealand Government funding, administered by the Royal
  Society of New Zealand. We thank the Kavli Institute for
  Theoretical Physics China at the Chinese Academy of Sciences for
  hospitality and support during the final stages of writing this
  article. Work at Argonne was supported by DOE BES under Contract
  No.\ DE-AC02-06CH11357. U.~Z.\ gratefully acknowledges hospitality
  at the Aspen Center for Physics during the 2008 Summer program.
\end{acknowledgments}


\begin{thebibliography}{10}

\bibitem{gei07}
A.~K. Geim and K.~S. Novoselov, Nature Mater. \textbf{6}, 183 (2007).

\bibitem{bee08}
C.~W.~J. Beenakker, Rev.\ Mod.\ Phys. \textbf{80}, 1337 (2008).

\bibitem{cas09}
A.~H. {Castro Neto}, F.~Guinea, N.~M.~R. Peres, K.~S. Novoselov, and A.~K.
  Geim, Rev.\ Mod.\ Phys. \textbf{81}, 109 (2009).

\bibitem{nov04}
K.~S. Novoselov, A.~K. Geim, S.~V. Morozov, D.~Jiang, Y.~Zhang, S.~V. Dubonos,
  I.~V. Grigorieva, and A.~A. Firsov, Science \textbf{306}, 666 (2004).

\bibitem{nov05b}
K.~S. Novoselov, D.~Jiang, F.~Schedin, T.~J. Booth, V.~V. Khotkevich, S.~V.
  Morozov, and A.~K. Geim, Proc. Natl. Acad. Sci. USA \textbf{102}, 10451
  (2005).

\bibitem{bir74}
G.~L. Bir and G.~E. Pikus, \emph{Symmetry and Strain-Induced Effects in
  Semiconductors} (Wiley, New York, 1974).

\bibitem{wal47}
P.~R. Wallace, Phys.\ Rev. \textbf{71}, 622 (1947).

\bibitem{sem84}
G.~W. Semenoff, Phys.\ Rev.\ Lett. \textbf{53}, 2449 (1984).

\bibitem{div84}
D.~P. DiVincenzo and E.~J. Mele, Phys.\ Rev.~B \textbf{29}, 1685 (1984).

\bibitem{hal88}
F.~D.~M. Haldane, Phys.\ Rev.\ Lett. \textbf{61}, 2015 (1988).

\bibitem{sai98}
R.~Saito, G.~Dresselhaus, and M.~S. Dresselhaus, \emph{Physical Properties of
  Carbon Nanotubes} (Imperial College, London, 1998).

\bibitem{pai70}
G.~S. Painter and D.~E. Ellis, Phys.\ Rev.~B \textbf{1}, 4747 (1970).

\bibitem{car53}
J.~L. {Carter Jr.}, Ph.D. thesis, Cornell University (1953).

\bibitem{lom55}
W.~M. Lomer, Proc. R. Soc. Lond. A \textbf{227}, 330 (1955).

\bibitem{slo55}
J.~C. Slonczewski, Ph.D. thesis, Rutgers University (1955).

\bibitem{slo58}
J.~C. Slonczewski and P.~R. Weiss, Phys.\ Rev. \textbf{109}, 272 (1958).

\bibitem{bas67}
F.~Bassani and G.~{Pastori Parravicini}, Nuovo Cim. B \textbf{50}, 95 (1967).

\bibitem{lut56}
J.~M. Luttinger, Phys.\ Rev. \textbf{102}, 1030 (1956).

\bibitem{win03}
R.~Winkler, \emph{Spin-Orbit Coupling Effects in Two-Dimen\-sion\-al Electron
  and Hole Systems} (Springer, Berlin, 2003).

\bibitem{man07}
J.~L. {Ma\~nes}, F.~Guinea, and M.~A.~H. Vozmediano, Phys.\ Rev.~B \textbf{75},
  155424 (2007).

\bibitem{and00}
T.~Ando, J.~Phys.\ Soc.\ Jpn. \textbf{69}, 1757 (2000).

\bibitem{mar02}
A.~{De Martino}, R.~Egger, K.~Hallberg, and C.~A. Balseiro, Phys.\ Rev.\ Lett.
  \textbf{88}, 206402 (2002).

\bibitem{kan05a}
C.~L. Kane and E.~J. Mele, Phys.\ Rev.\ Lett. \textbf{95}, 226801 (2005).

\bibitem{zar09}
M.~Zarea and N.~Sandler, Phys.\ Rev.~B \textbf{79}, 165442 (2009).

\bibitem{cha77}
D.~J. Chadi, Phys.\ Rev.~B \textbf{16}, 790 (1977).

\bibitem{subset}
If the IRs $\{\Gamma_\kappa\}$ contained in the product $\Gamma_\alpha \times
  \Gamma_\beta$ are only a subset of all IRs $\{\Gamma_{\kappa'}\}$ of
  $\mathcal{G}$, a tensor operator $\vekc{K}^{(\kappa',\lambda)}$ transforming
  according to an IR $\Gamma_{\kappa'} \notin \{\Gamma_\kappa\}$ is forbidden
  by symmetry to occur in $\mathcal{H}_{\alpha\beta} (\vekc{K}) $. For example,
  in $\mathcal{H}_{55}^\vek{K} (\vekc{K})$ all tensor operators transforming
  according to $\Gamma_3$, $\Gamma_4$, and $\Gamma_5$ are forbidden by
  symmetry. Yet these tensor operators are allowed in $\mathcal{H}_{77}^\vek{K}
  (\vekc{K})$, $\mathcal{H}_{99}^\vek{K} (\vekc{K})$, and/or
  $\mathcal{H}_{79}^\vek{K} (\vekc{K})$, see Table~\ref{tab:basemat}.

\bibitem{kos63}
G.~F. Koster, J.~O. Dimmock, R.~G. Wheeler, and H.~Statz, \emph{Properties of
  the Thirty-Two Point Groups} (MIT, Cambridge, MA, 1963).

\bibitem{koster_56}
We define the IRs of $D_{3h}$ via the characters listed in Table 65 of
  Ref.~\onlinecite{kos63}. This requires that the basis functions for
  $\Gamma_5$ and $\Gamma_6$ of $D_{3h}$ are reversed, see, e.g.,
  Ref.~\onlinecite{bir74}. To match the coordinate system in our
  Fig.~\ref{fig:lattice}, we have recalculated the coupling coefficients of
  $D_{3h}$ listed in Table~67 of Ref.~\onlinecite{kos63} for a coordinate
  system, where the in-plane symmetry axes are rotated by $30^\circ$ relative
  to those defined in Fig.~3 of Ref.~\onlinecite{kos63}.

\bibitem{sak85}
J.~J. Sakurai, \emph{Modern Quantum Mechanics} (Addison-Wesley, Redwood City,
  1985).

\bibitem{rau75}
H.~Rauch, A.~Zeilinger, G.~Badurek, and A.~Wilfing, Phys.\ Lett.~A \textbf{54},
  425 (1975).

\bibitem{wer75}
S.~A. Werner, R.~Colella, A.~W. Overhauser, and C.~F. Eagen, Phys.\ Rev.\ Lett.
  \textbf{35}, 1053 (1975).

\bibitem{sak67}
J.~J. Sakurai, \emph{Advanced Quantum Mechanics} (Addison-Wesley, Reading, MA,
  1967).

\bibitem{dre55a}
G.~Dresselhaus, Phys.\ Rev. \textbf{100}, 580 (1955).

\bibitem{byc84}
Y.~A. Bychkov and E.~I. Rashba, J.\ Phys.~C: Solid State Phys. \textbf{17},
  6039 (1984).

\bibitem{envelope}
For a Hamiltonian like Eq.\ (\ref{eq:invar}) or the Dirac Hamiltonian the full
  wave function is the product of the solution of this Hamiltonian times the
  corresponding basis functions.

\bibitem{dirac}
The (restricted) homogenous Lorentz group is often denoted $SO(3,1)$. To
  account for spin, we need its universal covering group, which is the special
  linear group $SL(2)$.

\bibitem{ras60}
E.~I. Rashba, Sov. Phys.--Solid State \textbf{2}, 1109 (1960).

\bibitem{wurtzite}
A similar situation occurs, e.g., at the $K$ points of wurtzite materials,
  \cite{bir74} yet in the absence of a center of inversion.

\bibitem{zub03}
K.~Zuber, \emph{Neutrino Physics} (Taylor and Francis, Bristol, 2003).

\bibitem{mcc56}
J.~W. McClure, Phys.\ Rev. \textbf{104}, 666 (1956).

\bibitem{lan7e}
L.~D. Landau and E.~M. Lifshitz, \emph{Theory of Elasticity} (Pergamon, Oxford,
  1987), 3rd ed.

\bibitem{tea09}
M.~L. Teague, A.~P. Lai, J.~Velasco, C.~R. Hughes, A.~D. Beyer, M.~W. Bockrath,
  C.~N. Lau, and N.-C. Yeh, Nano Lett. \textbf{9}, 2542 (2009).

\bibitem{bao09}
W.~Bao, F.~Miao, Z.~Chen, H.~Zhang, W.~Jang, C.~Dames, and C.~N. Lau, Nature
  Nanotech. \textbf{4}, 562 (2009).

\bibitem{suz02a}
H.~Suzuura and T.~Ando, Phys.\ Rev.~B \textbf{65}, 235412 (2002).

\bibitem{man07a}
J.~L. {Ma\~nes}, Phys.\ Rev.~B \textbf{76}, 045430 (2007).

\bibitem{kan97}
C.~L. Kane and E.~J. Mele, Phys.\ Rev.\ Lett. \textbf{78}, 1932 (1997).

\bibitem{voz10}
M.~A.~H. Vozmediano, M.~I. Katsnelson, and F.~Guinea, Phys. Rep. \textbf{496},
  109 (2010).

\bibitem{fog08}
M.~M. Fogler, F.~Guinea, and M.~I. Katsnelson, Phys.\ Rev.\ Lett. \textbf{101},
  226804 (2008).

\bibitem{per09}
V.~M. Pereira and A.~H. {Castro Neto}, Phys.\ Rev.\ Lett. \textbf{103}, 046801
  (2009).

\bibitem{gui10}
F.~Guinea, M.~I. Katsnelson, and A.~K. Geim, Nature Phys. \textbf{6}, 30
  (2010).

\bibitem{jua07}
F.~{de Juan}, A.~Cortijo, and M.~A.~H. Vozmediano, Phys.\ Rev.~B \textbf{76},
  165409 (2007).

\bibitem{par08}
C.-H. Park, L.~Yang, Y.-W. Son, M.~L. Cohen, and S.~G. Louie, Nature Phys.
  \textbf{4}, 213 (2008).

\bibitem{gui08}
F.~Guinea, B.~Horovitz, and P.~{Le Doussal}, Phys.\ Rev.~B \textbf{77}, 205421
  (2008).

\bibitem{dre65}
G.~Dresselhaus and M.~S. Dresselhaus, Phys.\ Rev. \textbf{140}, A401 (1965).

\bibitem{min06}
H.~Min, J.~E. Hill, N.~A. Sinitsyn, B.~R. Sahu, L.~Kleinman, and A.~H.
  MacDonald, Phys.\ Rev.~B \textbf{74}, 165310 (2006).

\bibitem{hue06}
D.~Huertas-Hernando, F.~Guinea, and A.~Brataas, Phys.\ Rev.~B \textbf{74},
  155426 (2006).

\bibitem{gmi09}
M.~Gmitra, S.~Konschuh, C.~Ertler, C.~Ambrosch-Draxl, and J.~Fabian, Phys.\
  Rev.~B \textbf{80}, 235431 (2009).

\bibitem{hue09}
D.~Huertas-Hernando, F.~Guinea, and A.~Brataas, Phys.\ Rev.\ Lett.
  \textbf{103}, 146801 (2009).

\bibitem{mal09}
L.~M. Malard, M.~H.~D. {Guimar\~aes}, D.~L. Mafra, M.~S.~C. Mazzoni, and
  A.~Jorio, Phys.\ Rev.~B \textbf{79}, 125426 (2009).

\bibitem{nov05a}
K.~S. Novoselov, A.~K. Geim, S.~V. Morozov, D.~Jiang, M.~I. Katsnelson, I.~V.
  Grigorieva, S.~V. Dubonos, and A.~A. Firsov, Nature \textbf{438}, 197 (2005).

\bibitem{zha05c}
Y.~Zhang, Y.-W. Tan, H.~L. Stormer, and P.~Kim, Nature \textbf{438}, 201
  (2005).

\bibitem{win07}
R.~Winkler, U.~Z\"ulicke, and J.~Bolte, Phys.\ Rev.~B \textbf{75}, 205314
  (2007).

\bibitem{cul08a}
D.~Culcer, Int. J. Mod. Phys. B \textbf{22}, 4765 (2008).

\end{thebibliography}
\end{document}